\documentclass[rmp, onecolumn, superscriptaddress]{revtex4-2}
\usepackage[utf8]{inputenc}
\usepackage[T1]{fontenc}
\usepackage{graphicx}
\usepackage[hidelinks, colorlinks=True, citecolor=blue ]{hyperref}
\setcitestyle{numbers,square}
\usepackage{caption}

\makeatletter
\let\cat@comma@active\@empty
\makeatother

\begin{document}

\title{Citizen science to assess light pollution with mobile phones}

\author{Gorka Mu\~noz-Gil}
\thanks{These authors contributed equally to the work}
\affiliation{ICFO -- Institut de Ci\`encies Fot\`oniques, The Barcelona Institute of Science and Technology, Av. Carl Friedrich Gauss 3, 08860 Castelldefels (Barcelona), Spain}
\affiliation{Institute for Theoretical Physics, University of Innsbruck, Technikerstr. 21a, A-6020 Innsbruck, Austria}

\author{Alexandre Dauphin}
\thanks{These authors contributed equally to the work}
\affiliation{ICFO -- Institut de Ci\`encies Fot\`oniques, The Barcelona Institute of Science and Technology, Av. Carl Friedrich Gauss 3, 08860 Castelldefels (Barcelona), Spain}

\author{Federica A. Beduini}
\email[Correspondence: ]{federica.beduini@icfo.eu}
\affiliation{ICFO -- Institut de Ci\`encies Fot\`oniques, The Barcelona Institute of Science and Technology, Av. Carl Friedrich Gauss 3, 08860 Castelldefels (Barcelona), Spain}

\author{Alejandro Sánchez de Miguel}
\affiliation{Depto. Física de la Tierra y Astrofísica, Instituto de Física de Partículas y del Cosmos (IPARCOS), Universidad Complutense, Madrid, Spain}
\affiliation{Environment and Sustainability Institute, University if Exeter, Penryn, Cornwall TR10 9FE, U.K.}
\affiliation{Instituto de Astrofísica de Andalucía, Glorieta de la Astronomía, s/n, 18008, Granada, Spain}

\begin{abstract}
The analysis of the colour of artificial lights at night has impact on diverse fields, but current data sources have either limited resolution or scarce availability of images for a specific region. In this work, we propose crowdsourced photos of streetlights as an alternative data source: for this, we designed NightUp Castelldefels, a pilot for a citizen-science experiment aimed at collecting data about the colour of streetlights. In particular, we extract the colour from the collected images and compare it to an official database, showing that it is possible to classify streetlights according to their colour from photos taken by untrained citizens with their own smartphones. We also compare our findings to the results obtained from one of the current sources for this kind of studies. The comparison highlights how the two approaches give complementary information about artificial lights at night in the area. This work opens a new avenue in the study of the colour of artificial lights at night with the possibility of accurate, massive and cheap data collection.
\end{abstract}

\maketitle

\section{Introduction}
\label{sec1}

Artificial lights at night are an important indicator of human activity and of their impact on the environment. Their study allows for an estimation of many complex global parameters, with applications ranging from economy - allowing for predictions of gross domestic product~\cite{forbes2013multi}, population density~\cite{ghosh2010shedding} and energy consumption~\cite{elvidge1997relation,sanchezde2014evolution} - to environmental sciences - helping in the estimation of carbon dioxide emissions~\cite{doll2000night,asefi2014multiyear}, sky brightness~\cite{de2020nature,falchi2016new}, and landscape connectivity~\cite{laforge2019reducing}. Most of the above-mentioned studies focus on light intensity, with data collected from panchromatic remote sensing instruments placed on satellites, such as the DMSP-OLS~\cite{elvidge1997relation} and the VIIRS-DNB from the Suomi North Polar Partnership (SNPP)~\cite{roman2018nasa}. 

However, important information can also be extracted from the spectra of light sources. Indeed, recent works have demonstrated that different ranges of the visible spectrum impact differently a plethora of biological processes, affecting profoundly the flora~\cite{lian2021artificial}, the fauna~\cite{foster2021light,pauwels2019accounting,cohnstaedt2008light} and human communities~\cite{chepesiuk2009missing,garcia2018evaluating}. In particular, according to various studies on ecological and health indicators~\cite{garcia2018evaluating, aube2013evaluating, gray2011reduced}, it is the blue range of the spectrum that affects most the living beings. 

Currently, there is not enough field surveys that relates color ambient light with health impacts. \cite{mcisaac2021impact} did a field survey to compare the intensity but not the resolution. And \cite{huss2019shedding} showed the limitations of comparing uncalibrated and low resolution ISS images with field data, again without color information. A summary of all health and environmental impacts ca be found on \cite{gaston2022environmental}.

At the same time, we have poor estimations of the light intensity in this range, due to the blindness to blue light of the instruments on satellites mentioned above. For example, current estimations of the growth of blue light global emission in the 1992-2017 period varies between 49\% to 270\% depending on the different scenarios. Even, recent publications show how European Union light emission have increased 13\% on green light and 25\% in blue light \cite{sanchez2022environmental}. It is thus evident that it is crucial to find other data sources to estimate the blue contribution to light emissions, even with limited accuracy, in order to reduce significantly this error and obtain better estimations of the radiance~\cite{kyba2017artificially,sanchez2021first} that would lead to more accurate economical and ecological indicators\cite{levin2020remote}.
It has been demonstrated that obtaining at least an approximation of the colour temperature of the light sources is important\cite{longcore2018rapid}: while it does not give as complete information as a spectrum, it is a sufficient parameter to determine the potential environmental impact.

Various systems have been tested to determine at least the colour temperature of the lighting technologies used in different regions of the world. Photo-metric techniques have limitations, either in spatial and spectral resolution, that make it difficult and expensive to obtain. Images are thus the main data source for this kind of analysis, no matter if they have been taken from airplanes, cars, the International Space Station (ISS) or other smaller artificial satellites~\cite{levin2020remote}. Regarding satellite images, colored data are currently available only from two sources: the photos taken by astronauts from the International Space Station (ISS) with a commercial digital single-lens reflex camera~\cite{de2019colour,de2021colour} and the ones generated by the commercial satellite JL1-03B~\cite{zheng2018new}. Nonetheless, both sources have serious limitations. For example, JL1-03B images are mainly accessible on request and cover few regions of the Earth. The ISS images coverage of the Earth surface is higher, but their resolution varies widely depending on the location. While low resolution images may be sufficient to represent large areas in some cases, they are not useful in densely populated regions, where outdoor lighting is usually extremely variable: this highlights the importance of collecting data with the highest possible resolution, especially in urban areas. Moreover, as most streetlamps emit light towards the ground, the spectra extracted from the images taken from satellites may be distorted, depending on the characteristics of the reflecting surfaces~\cite{shepherd2003correcting}.

It is also possible to infer blue light emissions from streetlight databases, which usually contain information about the location and the lighting technology (which determines the emission spectrum). However, it is difficult to obtain such data, as public lighting databases are usually managed by local authorities and their accessibility may vary depending on the local government. Moreover, an important contribution to light pollution comes from private illumination, for which no database exists. To circumvent these problems, researchers have used satellite images to assess and characterize light in the visible spectrum for specific regions. 

\begin{figure*}
\centering
\captionsetup{justification=centering}
    \includegraphics[width=0.7\columnwidth]{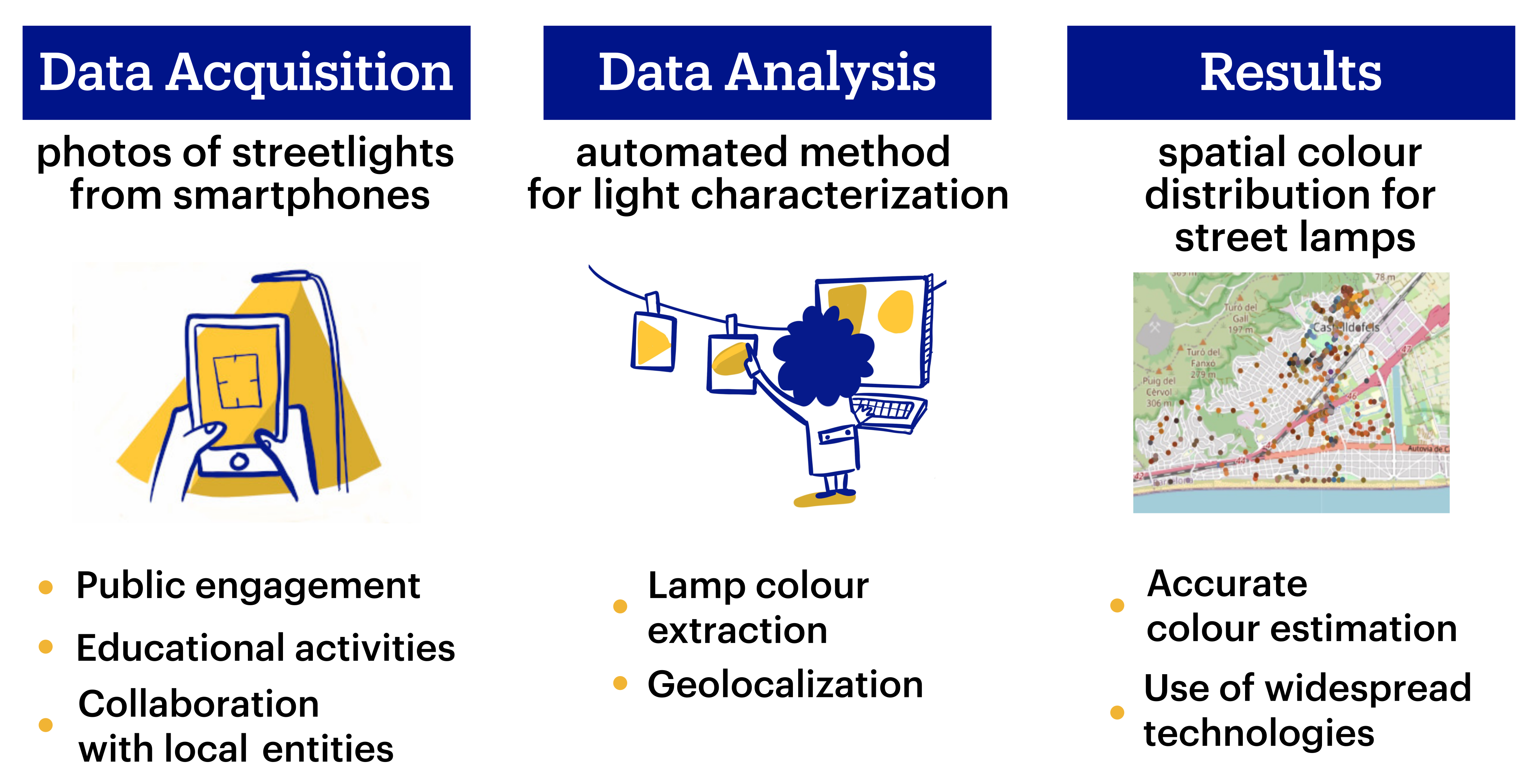}
	\caption{Schematic representation of the NightUp pipeline.}
	\label{fig_graph_abstract}

\end{figure*}

In this work, we propose the use of widely available light sensors - smartphone cameras - as a way to characterize the colour of artificial lights at night. This approach is a low-cost solution to complement and enrich satellite data while addressing some of the challenges of the other data sources. With people collecting photos of streetlights with their own smartphone, it is possible to collect information about the colour of artificial lights at night with unprecedented spatial resolution, as well as to reach urban areas that are not covered by currently available data sets. Furthermore, as the photos aim directly at the source of the light, they are not affected by the colour aberrations that affect satellite images due to reflection. To this aim, we have designed a citizen science experiment (NightUp, see Figure~\ref{fig_graph_abstract}), in which participants photograph artificial lights with their mobile phones. In this work, we analyse the data collected during the pilot NightUp campaign in the city of Castelldefels (Barcelona, Spain), showing the potential of mobile phones and citizen participation as an important source of data to study artificial lights at night.

\section{Materials and Methods}

\subsection{Data acquisition by citizen science}

Even though there are many definitions of citizen science~\cite{Haklay2021}, they converge on the fact that it involves crowdsourcing part of a scientific experiment (e.g. data collection, pattern recognition, data analysis, ...) to voluntary citizens outside the academic environment. This practice is very valuable, because it strengthens the relationship between the public and the academic world and transforms citizens from passive recipients of dissemination activities into active elements of the research process. At the same time, citizen science experiments are a very relevant tool for the scientific community, that can lead to solid scientific results that would be difficult without a collective effort due, for example, to the huge amount of data to be collected and/or classified~\cite{cooper2010predicting,lintott2011galaxy,big2018challenging,sherson2021quantummoves}. 

In the last twenty years, the advent of smartphones has become a game changer in citizen science. Indeed, their collections of sensors (microphones, location sensors, cameras, etc.) allows citizens to have a small laboratory in their pockets~\cite{land2016citizen,palmer2017citizen, odenwald2020smartphone}. 
More importantly, scientists now can potentially reach every corner of the world for their data collection campaigns. With this in mind, we have designed NightUp, a citizen science experiment that uses these widespread cameras to collect data missing in the light pollution scientific community: the spatial distribution of colour for artificial light at night. 

The success of a citizen science experiment like NightUp depends on two main factors: effective engagement strategies to gather and retain volunteers and a simple but rigorous method to collect the data to ensure that the citizen science practitioners deliver good quality data with minimum training. For the pilot phase, we have focused mainly on the latter factor, developing a mobile phone application (app) with an intuitive user experience and visual tutorials, allowing participants to collect data without special equipment, nor specific training. The app has been designed with a single and simple task for the participants: they only had to point their smartphones at the streetlights and take a photo through the app. The app would then upload it automatically to a server together with the geolocation of the device, its accuracy and a timestamp. Users could also answer optional questions about the light direction and the distance between similar lamps. The app collected the data anonymously, assigning an alphanumeric ID to each device, without gathering any other personal information.

\begin{figure*}
	\includegraphics[width=0.5\columnwidth]{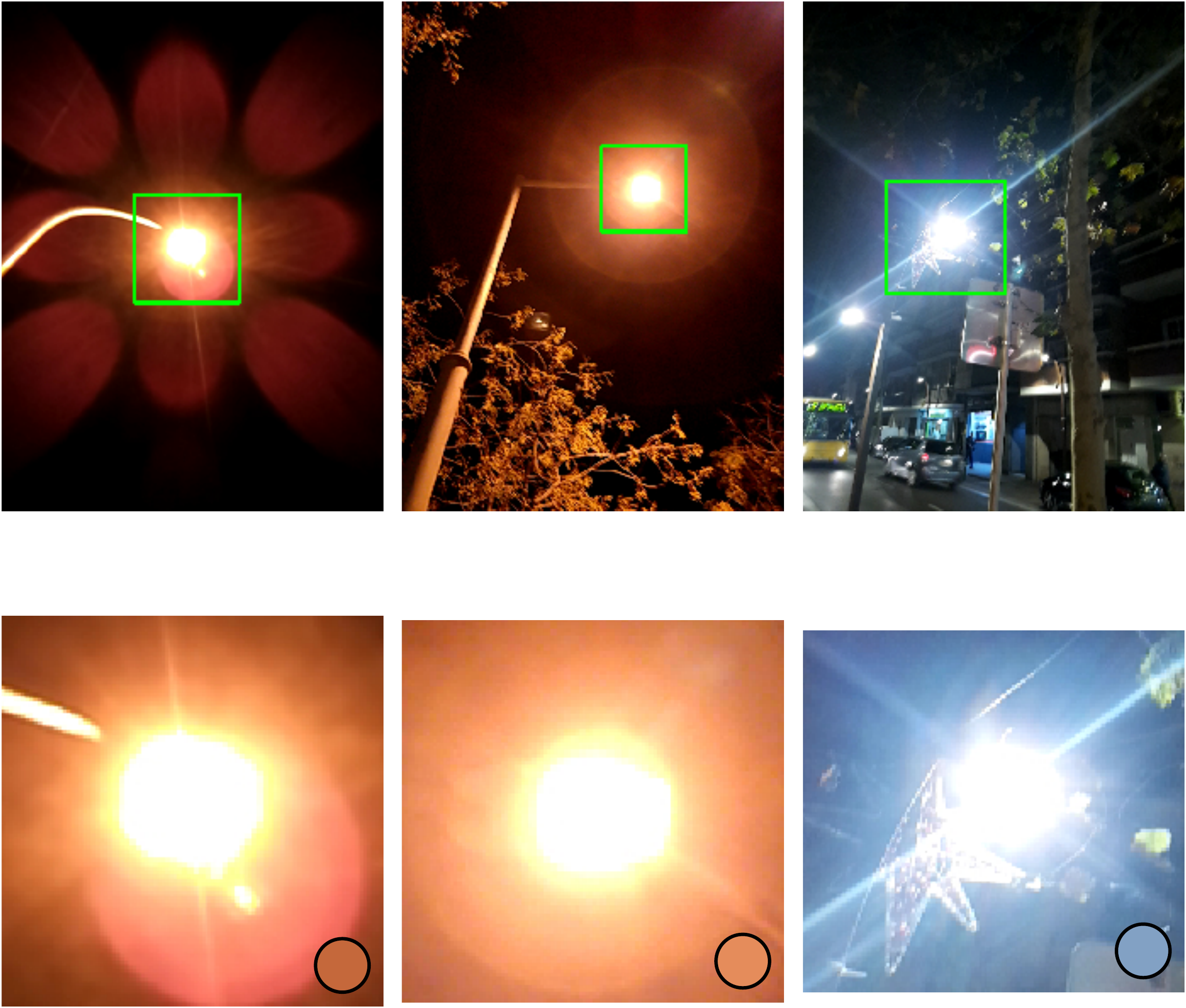}
	\caption{Top row: Three examples of light source location by means of our algorithm. Bottom row: The area in the rectangle of the image above is cropped and the circle in the bottom right corner shows the colour assigned to each lamp.}
	\label{fig:crop_box}
\end{figure*}

The idea of the NightUp method is that any person with a smartphone around the world should be able to contribute to the experiment with few simple steps. In order to compare results, the devices must be calibrated, but it is unrealistic to assume that each user could calibrate their own device or that we could do it for each model available on the market. We thus assumed that the calibration that smartphone cameras undergo to meet the ISO standards would be sufficient to obtain satisfactory results. Moreover, the app set the white balance of the smartphone camera to a colour temperature of 3000~K, when allowed by the device.

The data presented in this paper comes from the NightUp pilot campaign (NightUp Castelldefels), that was designed to test the hypothesis that no further calibration is needed to distinguish warmer light sources from colder ones. The pilot campaign lasted from November 2019 to March 2020, involving 74 unique users that contributed with 1112 photos. This first campaign was limited to the Castelldefels area, a mid-sized city that shows a variety of natural (beach, hills, plain) and artificial (industrial, commercial and residential areas) landscapes. Additional data from Castelldefels, Barcelona and El Prat de Llobregat areas were added in Fall 2020 and 2021, reaching a total of 1372 contributions. 

\subsection{Image analysis}
\label{sec:image_analysis}

In this section, we describe the three steps for the analysis of the photos taken with the NightUp application. The first step consists in manually cleaning the dataset from spurious images. The goal of the second step is to extract the colour of a given streetlight. To this end, we develop an algorithm to locate the region where the lamp is in the image. Then, we extract the average pixel colour to identify the colour temperature of the light. Finally, we associate all the extracted colours to their corresponding geolocalizations and plot their spatial distribution on a map of Castelldefels.  

\subsubsection{Manual verification}
\label{sec:manual}
In citizen science experiments, we often face the existence of incorrect samples, due for example to possible miscommunication between organizers and participants. This often generates invalid data, which potentially lead to inaccurate results. In our case, before starting the analysis process on the data set of the pilot campaign (November 2019-March 2020), we discarded the following types of photos (number and percentage w.r.t. the total pilot dataset in parenthesis): i) photos where the background is not completely dark (e.g. taken at dusk) that can lead to an overestimation of the blue component in our analysis (235 photos, 21.1 \%); ii) photos that - despite the instructions given - are not taken pointing directly at a light source (113 photos, 10.2 \%); iii) photos which are not streetlights (409 photos, 36.8 \%). Note that a photo can fail on various of the previous points.

After the manual verification, we have been left with 698 photos, a 62.8\% of the total photos taken by untrained citizens. The photos taken after March 2020, used in Section~\ref{sec:unlabelled}, were taken by users personally trained by the organizers, so that none of their photos have been discarded during the manual validation process.

\subsubsection{Colour extraction}
\label{sec:colour}

In order to find the light source in the uploaded photos and extract their colours, we developed an algorithm that works as follows: i) the image is transformed into its gray-scaled version; ii) we apply a Gaussian blur, a common procedure in image analysis to filter noisy images and enhance their segmentation~\cite{gedraite2011investigation}; iii) we set an intensity threshold to highlight only the brightest regions of the photos; iv) we select the largest region arising from the threshold, assuming that it corresponds to the lamp. It is a fair assumption, as we instructed citizens to point directly at the light source when shooting the photo.

\begin{figure*}
	\centering\includegraphics[width=0.6\columnwidth]{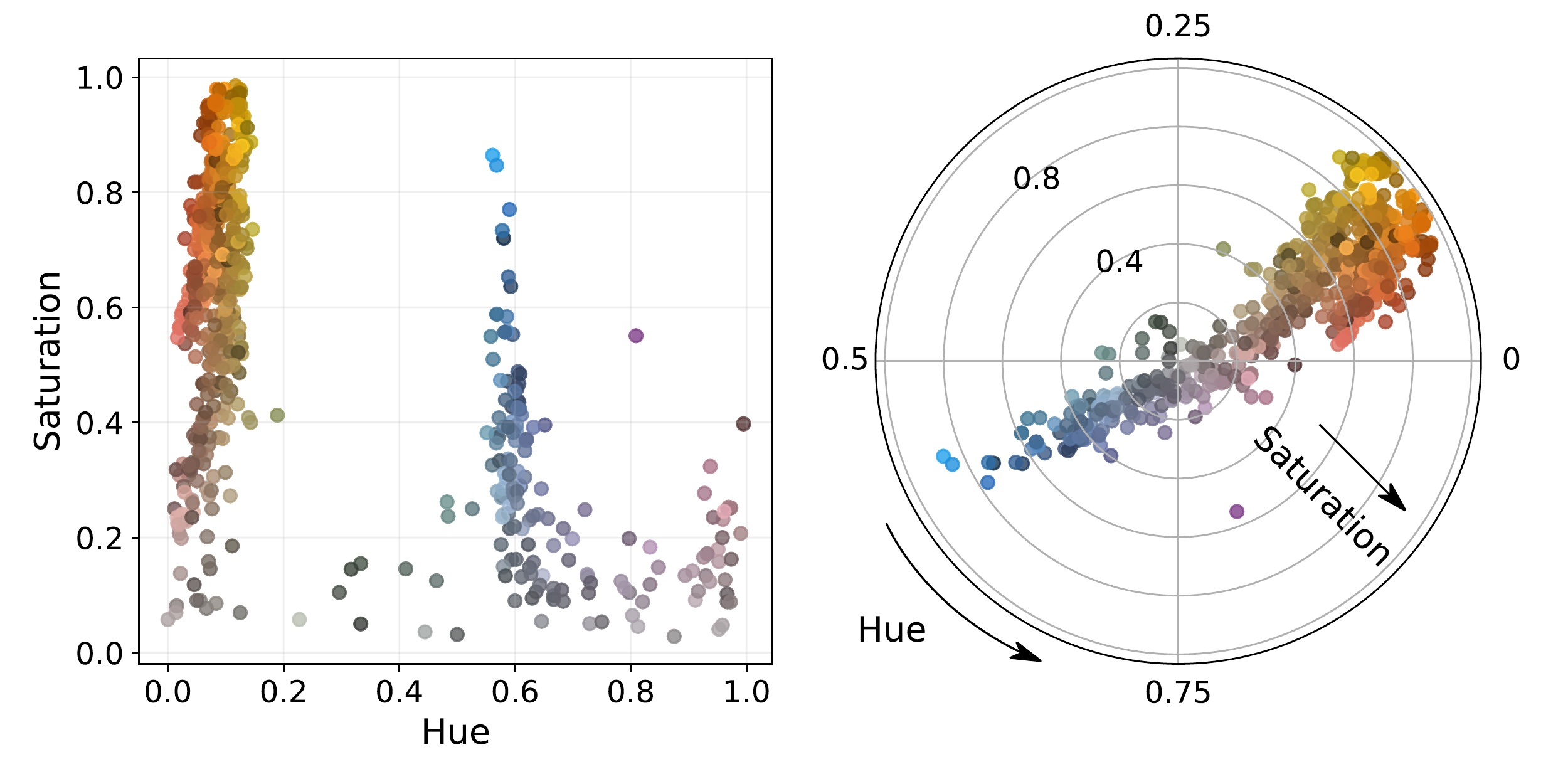}
	\caption{Scatter plot of the colour extracted from the NightUp photos as a function of the Saturation and Hue value, in cartesian (left) and polar (right) coordinates. The colour of each point is calculated with the procedure presented in Sec.~\ref{sec:colour}.}
	\label{fig:scatter}
\end{figure*}

We crop the region where the lamp is with a rectangular box large enough to contain both the light source and its aura. In fact, the lamp pixels are often saturated, so the aura - which shares the spectral properties of the lamp - is especially important to extract the colour information. To do this, we exclude all pixels in the cropped box that are saturated in at least one of the Red~(R), Green~(G) and Blue~(B) channels of the digital image. We average the value in each channel over the remaining pixels, obtaining the average colour of the box, which we assign to each image. Figure~\ref{fig:crop_box} shows the results of this procedure for three different photos, where each image has an assigned colour, depicted in the coloured circles.

Human experts have checked manually that our algorithm extracts the correct colour from all the photos (698) that passed the manual verification described in \ref{sec:manual}, proving our algorithm as a strong and robust method. If we apply the same procedure to the whole dataset (1389 photos), the segmentation was correct in 87.97\% of the photos, while the colour was correctly extracted in 79.4\%. 
This percentage could be improved in future campaigns, for example adding to the app automatic quality checks and feedback for the user before the submission of the photos. Moreover, the photos processed here may now serve as a valid training dataset for upcoming works, in which machine-learning-based segmentation algorithms~\cite{ronneberger2015u} may enhance the accuracy of the analysis of the photos.

For the following analysis, we consider only manually verified photos, which have been segmented correctly. We perform a change of basis from RGB to the Hue~(H), Saturation~(S) and Value~(V) model~\cite{burger2009}, which is particularly suitable to easily distinguish between blue and red. Figure~\ref{fig:scatter} shows the extracted colours of the lights in a H-S diagram. For our discussion, we focus on the H and S variables, because the value of V (associated to lightness) gives no information relative to the light spectrum. Notably, we see that two main regions arise, one with $0\leq H \leq 0.2$, corresponding to warm colours, and a second region with Hue values around $0.6$, corresponding to cold ones.

\begin{figure*}
	\centering\includegraphics[width=0.7\columnwidth]{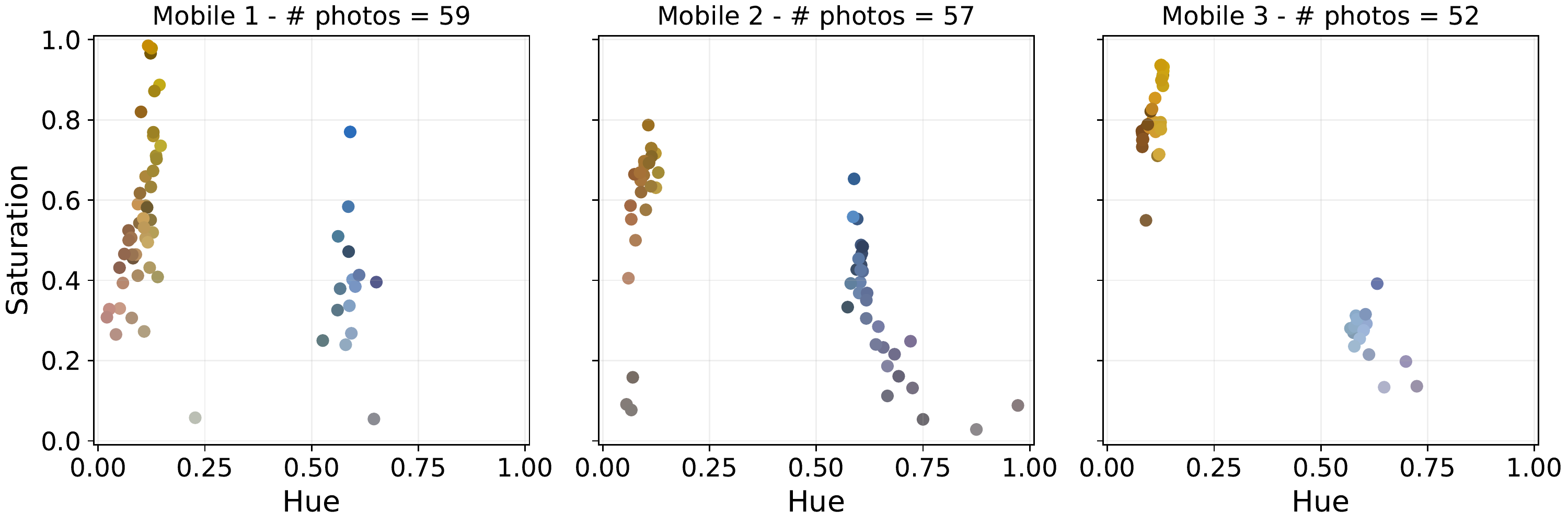}
	\caption{Scatter plots of the colour extracted from the photos taken by the three devices with more uploads to the NightUp application.}
	\label{fig:single_mobile}
\end{figure*}

An important assumption for the NightUp project is that smartphones cameras can successfully distinguish between different colours, even without additional calibration. To test the validity of such statement, we choose the three devices that took the most photos (59, 57 and 52, respectively) and plot their H-S diagram (see Figure~\ref{fig:single_mobile}). We observe that each device reproduces the separation of collected data into two main regions, as in  Figure~\ref{fig:scatter}, even if with less statistics, as expected.

\subsubsection{Spatial distribution of the colour of artificial light at night}

\begin{figure}
	\centering\includegraphics[width=0.8\columnwidth]{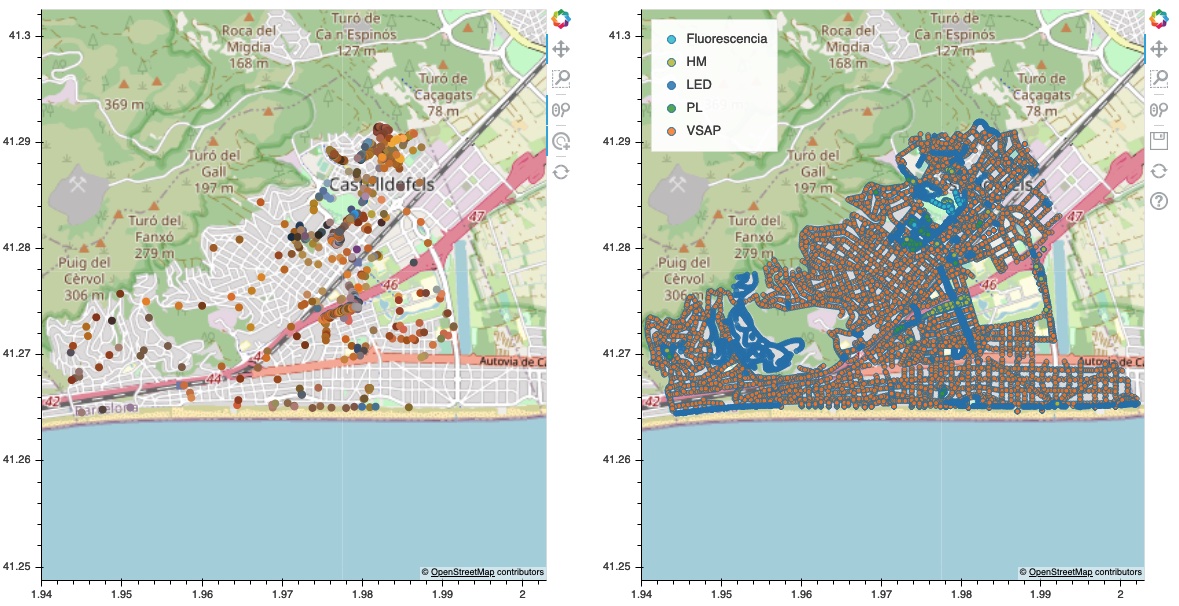}
	\caption{\textbf{(a)} Spatial distribution of the NightUp dataset for Castelldefels. \textbf{(b)} Spatial distribution of the streetlights as depicted in the dataset from the city council.}
	\label{fig:street_map}
\end{figure}

The NightUp app records the location of the smartphone when the photo is taken, so that we can represent the collected data on a map as in Figure~\ref{fig:street_map}(a), obtaining information about the spatial distribution of the colour of artificial light at night in Castelldefels. We observe that the data points are not homogeneously distributed: they appear mainly concentrated around the commercial area. This shows the importance of the collaboration with local entities (such as city council, library and schools) and of focused engagement campaigns for the data collection in order to cover wider territories. For instance, a data collection activity was organized in collaboration with the school \emph{Col·legi Frangoal} (Latitude: 41.288$^{\circ}$, Longitude: 1.984$^{\circ}$), resulting in almost 30 new devices and more than 580 photos in one night, increasing notably the density of photos in the north east area and along the central part of the northern highway (upper red road), as depicted in Figure~\ref{fig:street_map}(a).

Maps like the ones in Figure~\ref{fig:street_map}(a) have the potential to help the light pollution scientific community to estimate the impact of the colour of light in living beings and the blue contribution to global light emission.

\section{Results}

The NightUp dataset is composed of images taken with different devices with no specific calibration: it is therefore important to show that the colours extracted with the method described in Section~\ref{sec:image_analysis} provide valid information for scientific studies. To this aim, in this section, we compare our results with the streetlight database from the city council of Castelldefels, identifying regions in the H-S space that correspond to different colour categories that can be used for the characterization of unlabelled databases. Moreover, we compare the colours extracted from NightUp data to the ones derived from an image shot from the ISS, which is currently used as one of the main sources for this type of information.

\subsection{Comparing NightUp data to the city council streetlight database}
\label{sec:Castelldefels}
Thanks to the collaboration of Castelldefels city council in the NightUp project, we have access to a list of verified streetlights [see  Figure~\ref{fig:street_map}(b)], including their spatial coordinates and the lamp type.  This information allows us to associate each image with a potential real streetlight and validate the colour information extracted from the photo. It is important to note that such information might not be easy to obtain for many cities or regions: for this reason, an experiment like NightUp, once validated, could be an important source of information for the light pollution scientific community. Moreover, city councils' databases only cover public lightning managed by the municipality, thus not including private lights and the ones who might be managed by other kinds of local government.

\subsection{Geolocalization}
\label{sec:geolocalization}
The city council database distinguishes between five different lamp types, based on the technology used for light generation, with huge disparity in number (percentage in parenthesis):  high-pressure Sodium (HPS, $76.0 \%$), light emitting diode (LED, $17.4 \%$), plug-in fluorescent (PL, $2.4 \%$), metal-halide (MH, $2.4 \%$) and fluorescent (F, $1.8 \%$) lamps. The reported colour temperature for the different technologies are 3000-3500~K for LED, 4000~K for PL, 3500-4000~K for MH and 3500~K for F lamps.

Associating an image from the NightUp dataset to a lamp from the city council database is not trivial in some cases. In fact, the locations in our dataset have limited precision (15~m on average) and accuracy (users may have shot the photos from a certain distance of the streetlight). This may lead to mistakes in associating a NightUp image to a streetlight from the city council database, especially in areas with high density of streetlights of different types. To minimize such error source, we restrict our analysis to the photos for which the three closest lamps are of the same type, reducing to a $96.3\%$ (672 images) of the original, human-verified dataset.

\subsection{Colour extraction with different devices}
\begin{figure}[t]
	\centering\includegraphics[width=\columnwidth]{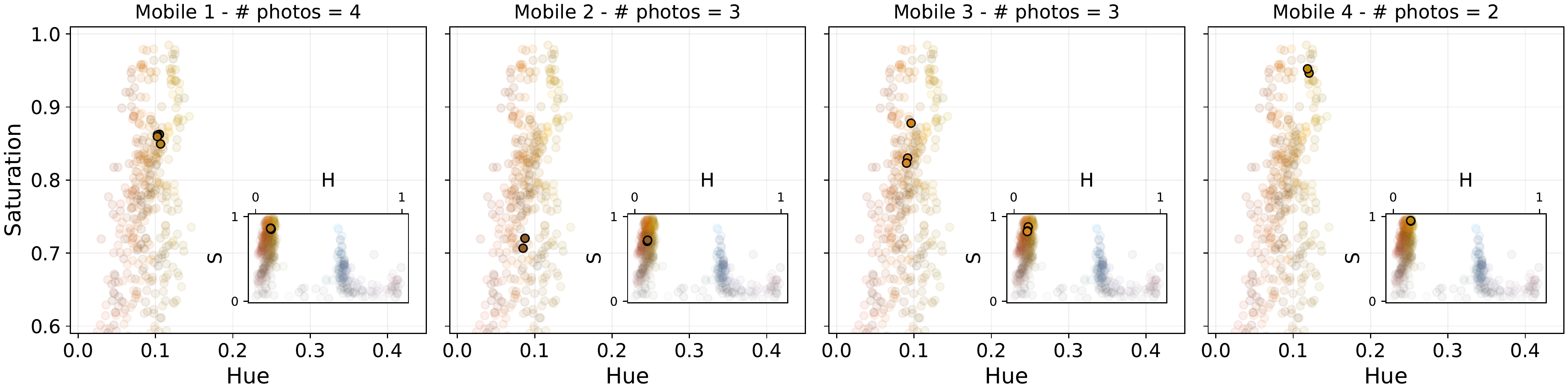}
	\caption{Scatter plots for the colours extracted from a set of photos taken in an area where all the lamps were verified to be of the same type (HPS). Each panel corresponds to the photos taken by a different device. The shaded points are the colours extracted from the photos of the whole human-verified database, for comparison. }
	\label{fig:single_lamp}
\end{figure}

In order to check whether smartphone cameras are able to provide consistent information about the colour of a light source, we investigate the Hue-Saturation distribution of the extracted colours from photos taken in a small area (Latitude $ \in [ 41.2884, 41.28882]$, Longitude $\in [1.9799, 1.981]$) in which only HPS lamps are present, according to the city council database. The assumption is that in a small area the lamps would be as similar as possible, allowing us to compare how different smartphone cameras respond to the same light source even without prior calibration. In Figure~\ref{fig:single_lamp} we highlight the colours corresponding to the photos taken by four different devices in that area. Even though the different points are not exactly in the same region of the H-S diagram, showing some dispersion among the different devices, all photos cluster in the top left of the diagram, corresponding to the warm colours associated with HPS lamps. This certifies that it is possible to identify colours within a certain precision with the NightUp method, even with different devices that have not been calibrated for this task.

\subsubsection{Lamp colour clusterization and automatic labelling}
\label{sec:lamp_type_clusters}
The previous figure suggests that - even though different devices may assign slightly different colours to the same light source - the colours extracted with our procedure for the same kind of lamp will cluster in the same region of the Hue-Saturation space.

\begin{figure}[t]
	\centering\includegraphics[width=0.55\columnwidth]{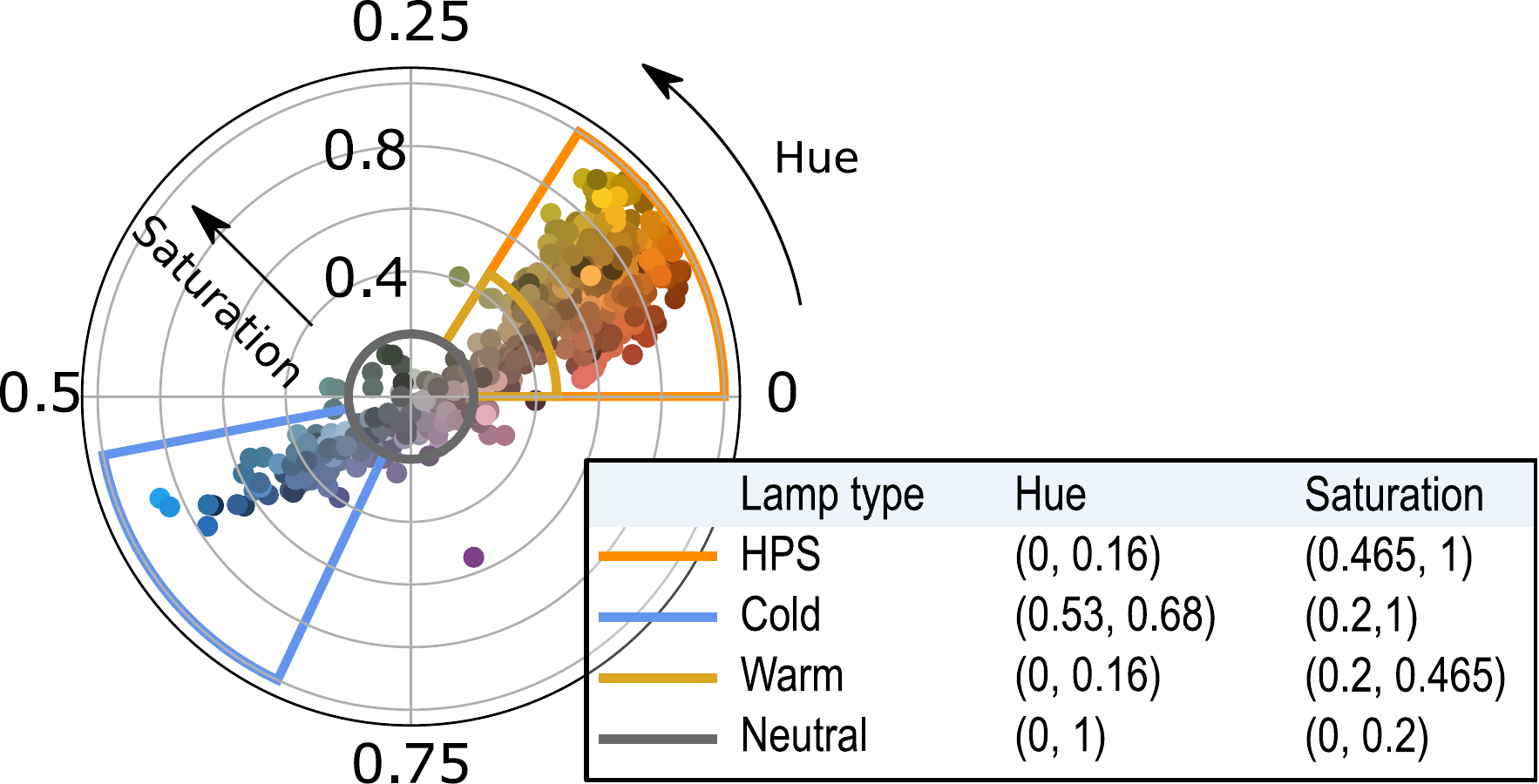}
	\caption{Scatter plot of the whole NightUp database as a function of saturation (distance from the center) and hue (angle). The coloured lines separate the regions of the H-S pace corresponding to different blue-light intensity levels. The boundaries of these regions are indicated in the legend.}
	\label{fig:regions}
\end{figure}

In fact, thanks to the comparison of our human-verified database with the city council, we observe that the majority of the extracted colours associated to HPS lamps via the method described in paragraph~\ref{sec:geolocalization} fall in the region marked with the orange border in Figure~\ref{fig:regions}. This region contains $82\%$ of the photos associated to HPS lamps and only $3.2\%$ of non-HPS lamps. For this reason, we can assume that a future photo falling in this region will be most likely a lamp that emits warm light, like a HPS one or a LED with color temperature smaller than 3000~K. 

Using only the average colour as information, it is difficult to distinguish between different light technologies that emit light in overlapping regions of the visible spectrum~\cite{de2019colour}. For this reason, we cannot assign a specific light technology to an area of the H-S diagram as we did for the HPS lamps.
Of course, having access to the lighting technology (and hence to the exact light spectrum) would be optimal. However, in many light pollution studies it is sufficient to be able to distinguish in which region of the light spectrum the lamp emits more light. For this reason, we focus on demonstrating that the NightUp method can successfully distinguish warm lamps, such as the HPS ones, with almost no emissions in the blue region of the spectrum, from other light sources with colour temperature higher then 3000~K that show more emissions in the blue region of the spectrum.

We decide to create three separated regions, as shown in Figure~\ref{fig:regions} assigned to cold (blue box, $33.3 \%$ of total non-HPS lamps), neutral (gray box, $33.3\%$) and warm (yellow box, $25 \%$) to classify the non-HPS lamps according to their amount of blue light in their average colour. The last region contains all the rest of the H-S space, that contains only a few outliers, usually related to poor quality images, in which phones wrongly capture the colours.

With this classification, it is easy to assign a new unlabelled photo to one of these categories, making it easy to distinguish between HPS lamps and the other sources of light and to characterize the light source according to its blue light content..

\subsection{Comparing NightUp data to ISS images}
As a further validation of our method, we compare our results to the ones obtained from the analysis of high resolution images taken by astronauts from the International Space Station (ISS).

\subsubsection{ISS image analysis}
\label{sec:ISS_sub}
We focus our study on the image ISS062-E-102578, which is the sharpest image of the area taken from the ISS in the period of the NightUp experiment. It is important to note that the ISS images are radiance calibrated~\cite{de2021colour}, and their pixel value is given in radiance units (nW$\cdot$ cm$^{-2}\cdot$ sr$^{-1}\cdot \hat{\mbox{A}}^{-1}$) (where nW refer to nanoWatts, cm to centimeters, sr to steradians and $\hat{\mbox{A}}$ to angstroms). On the other hand, the NightUp images are displayed in JPEG format, which means that their pixel values have arbitrary units calibrated according to the ISO standards~\cite{iso2006photography} and the correlated colour temperature (CCT) definition~\cite{judd1935maxwell}. In order to be able to compare the images from the two different sources, we apply a sequence of manipulations to the ISS image which effectively generate pixel values comparable to those of a JPEG images (see Appendix \label{app:iss_transformation} for the detailed procedure).

\subsubsection{Comparison between the NightUp and ISS results for the city of Casteldefells}

\begin{figure}
	\centering\includegraphics[width=\columnwidth]{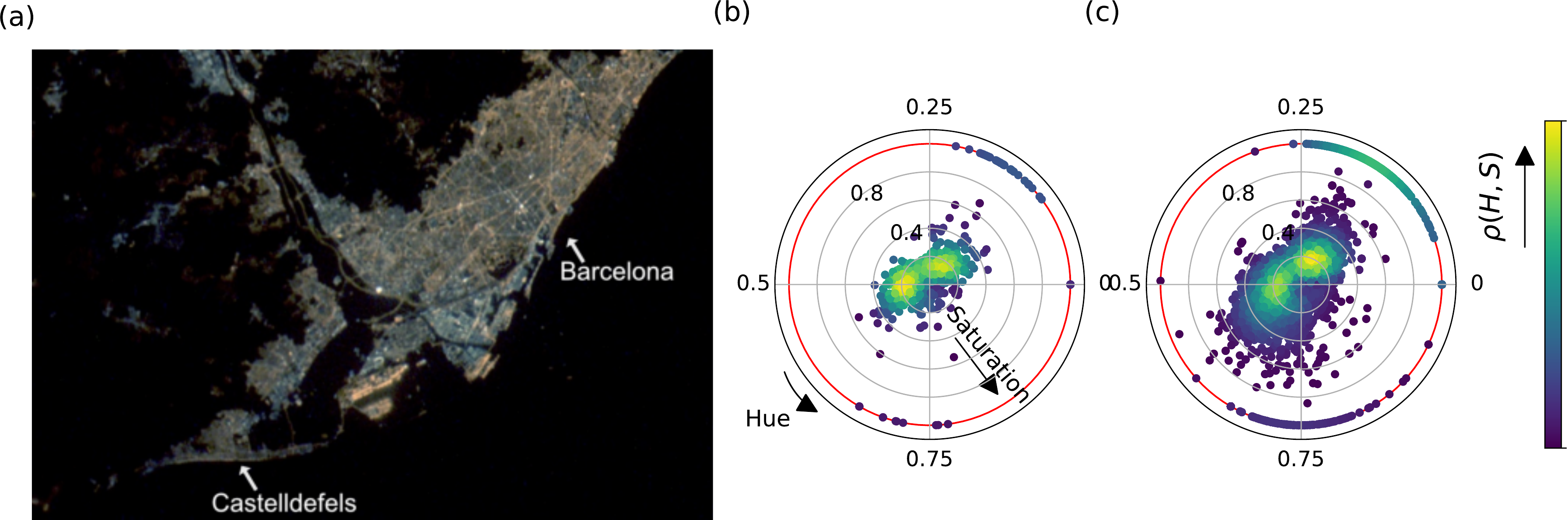}
	\caption{\textbf{(a)} Satellite image of the metropolitan area of Barcelona from the International Space Station. \textbf{(b)} H-S diagram of the colours of the pixels of picture (a) associated to the locations of NightUp photos. \textbf{(c)} H-S diagram of the colours of the pixels of picture (a) associated to the streetlights from the city council database. The colour of each point represents its kernel density estimation value w.r.t. a Gaussian kernel $\rho(H,S)$ (see main text for details). The outer red rings in picture (b) and (c) correspond to the saturation value $S=1$.} 
	\label{fig:iss}
\end{figure}

Figure~\ref{fig:iss}(a) depicts the ISS image ISS062-E-102578 centered in the Barcelona area, after the transformation described in the previous paragraph. In this image, we identify the pixels corresponding to the geolocations given by: i) the Nightup photos; ii) the Castelldefels city council database. We then extract the colour for each one of these pixels and plot the corresponding H-S values in Figure~\ref{fig:iss}(b-c) for each set of localizations. As opposed to previous plots, the colour of each point here represents the kernel density estimation w.r.t. to a Gaussian kernel, $\rho(H,S)$~\cite{hastie2009elements}. As the number of data points is much larger than in previous cases, with this procedure we can show an accurate approximation of the density distribution of lamps in the H-S space. Here, in contrast to the NightUp analysis (see Figure~\ref{fig:scatter}), the distribution of colours is heavily concentrated around values of H$\sim 0$ and does not yield a clear distinction between different lamp colours. 

It is important to note that the ISS images have much lower \textit{spatial} resolution than the NightUp data. For instance, the ISS sampling is 25 meters per pixel, with a Point Spread Function (PSF) of 75 meters. Moreover, the ISS image include \textit{all} sources of artificial light at night, including the ones that are not public streetlights. While the NightUp dataset also contains non-streetlights sources, the users were asked to make photos \textit{only} from the street, hence lowering the contribution of other light sources (e.g. lights in private gardens or soccer fields). Another important source of error is the sensitivity towards obstacles for the ISS images: for example,  the ISS cannot detect the streetlights under dense canopies of trees. Moreover, streetlights are usually directed towards the ground, as their main objective is illuminating the streets. In this case, the ISS images can capture their reflection, which affects the measured colours~\cite{shepherd2003correcting} and in some cases the direct light. On the other hand, NightUp photos capture the light source directly, without obstacles or reflections that may change the measured colour.

Finally, we observe in both Figure \ref{fig:iss}(b-c) that many points show saturation values $S=1$ (outer ring in red). All these points correspond to colours with very low value V, meaning that the associated pixels were mainly dark (black). This could arise mostly from two phenomena: i) the presence of an obstacle between the light emitter and the satellite; ii) the ISS images are not automatically assigned to a location, but they are georeferenced with the procedure described in \cite{de2021colour}; eventual errors in georeferencing the ISS images lead to extract the colour of pixel where no lamp is present. These two effects are not present in NightUp, as i) the photos are directly aimed at the light source, and hence no obstacle is expected; ii) the geolocation is automatically assigned to the photo by the NightUp app, even if with some smaller errors, as shown in Section~\ref{sec:geolocalization} and Figure~\ref{fig:street_map}.

\begin{figure}[b]
	\centering\includegraphics[width=\columnwidth]{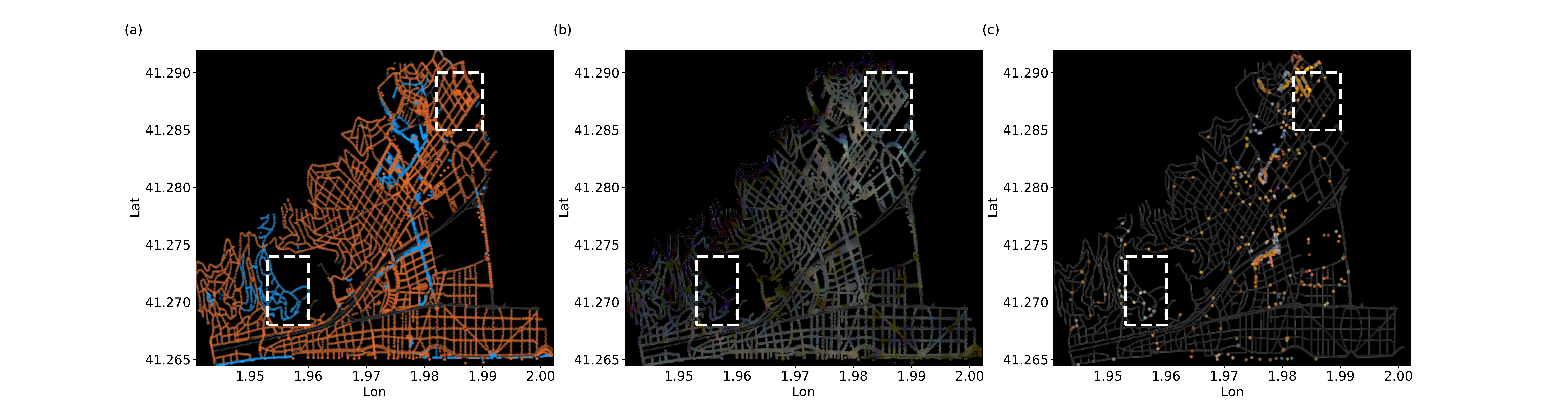}
	\caption{\textbf{(a)} Inventory of the streetlights of Castelldefels, distinguishing the HPS lamps (orange) from the other types (blue) \textbf{(b)} Colours of the streetlights according to the ISS data. \textbf{(c)} Colours of the streetlights according to the NightUp data (where available). Selected areas are marked by the white dashed line.} 
	\label{fig:comparison_nup_iss_ch}
\end{figure}

In Figure~\ref{fig:comparison_nup_iss_ch} we show the map of Castelldefels and highlight: a) the lamps of the city council database, as well as their type (orange for HPS, blue for the rest); b) the colour of the pixels associated to the city council database lamps as extracted from the ISS image; c) the colour of the images of the NightUp dataset. This figure highlights some differences between the information that can be extracted from the two sources. For instance, the ISS image allows for a complete characterization of all the lamps in the municipality with just one image. However, it is difficult to distinguish regions with different lamp types, as we have shown in Figure~\ref{fig:iss}. On the contrary, the NightUp approach allows for a much better characterization of the colour, allowing us to distinguish HPS lamps from the rest. However, data are not uniformly distributed and mainly concentrated in the most populated areas. Nevertheless, this can be addressed with more focused advertising and the collaboration with local entities in future NightUp campaigns. We note that the data presented in this paper was acquired at different times: the NightUp dataset was gathered from November 2019 to March 2020, the city council's database was issued in Summer 2020 while the ISS image dates of March 2020. While some changes to the public streetlights may have happened in between, those are minimal and we don't expect them to affect our analysis.

\begin{figure*}
	\centering\includegraphics[width=\columnwidth]{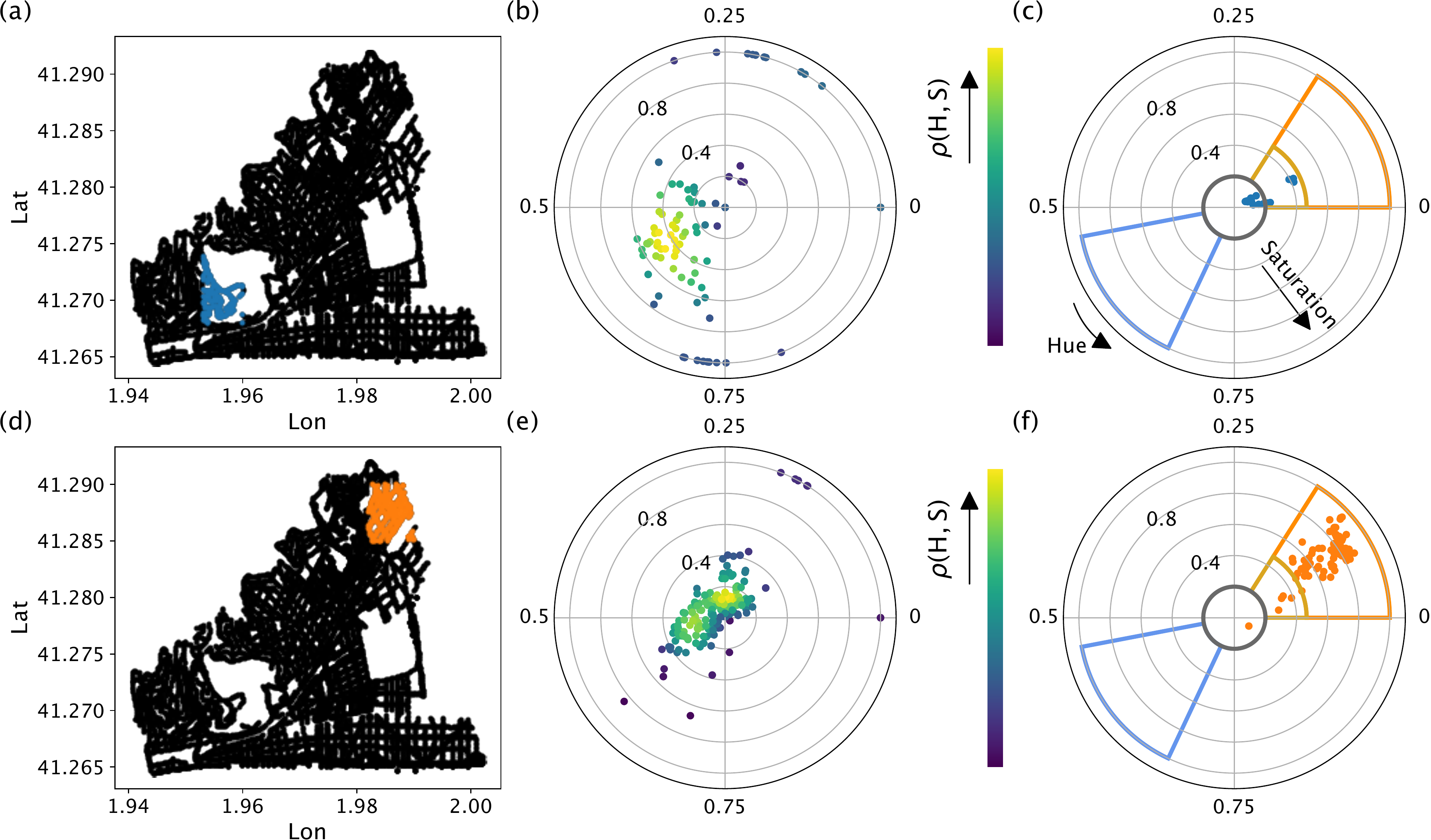}
	\caption{\textbf{Top row:} Comparison between the colours from the ISS (b) and from NightUp (c) for a region labelled by the city council of Casteldefells as LEDs (a). \textbf{Bottom row:} Comparison between the colours from the ISS (e) and from NightUp (f) for a region labelled by the city council of Casteldefells as HPS (d). The colour of each point in (b,e) represent its kernel density estimation value w.r.t. a Gaussian kernel $\rho(H,S)$.} 
	\label{fig:comp_iss_nup}
\end{figure*}

To further extend our comparison, we focus now on two areas of Castelldefels: each one contains lamps from the same lamp category as given by the city council database. Both are marked in Figure~\ref{fig:comparison_nup_iss_ch} with white boxes and further highlighted in Figure~\ref{fig:comp_iss_nup}. The top-right area of the maps contains only HPS lamps, while the bottom-left contains only LED lamps. The extracted colours from the ISS image and NightUp are shown in Figure~\ref{fig:comp_iss_nup}(b,e) and (c,f), respectively. We first focus on the LED region (Figure~\ref{fig:comp_iss_nup}(a-c)), where 24 photos where acquired with NightUp and 174 points extracted from the ISS image. All the NightUp points cluster in the Neutral and Warm regions, which in both cases refer to non-HPS images, as shown in Section~\ref{sec:lamp_type_clusters} and Figure~\ref{fig:regions}. On the other hand, the colours extracted from the ISS show a huge dispersion, without a clear clusterization. In the HPS area, 86 points in H-S diagram come from NightUp and 411 from the ISS. Importantly, $94.2 \%$ of the NightUp photos are located in the HPS colour region (Figure~\ref{fig:comp_iss_nup}(f)). The colours extracted from the ISS showcase here more a pronounced cluster, compared to plot the LED area. However, the colours extracted from the ISS have low V values and concentrate in similar regions of the H-S diagram for the LED and HPS areas, making it difficult to distinguish between lamp categories. This further proves that the NightUp approach is an important tool for colour characterization in artificial lights at night.

\subsubsection{Comparison between the NightUp and ISS results on an unlabelled dataset}
\label{sec:unlabelled}

\begin{figure}
	\centering\includegraphics[width=\columnwidth]{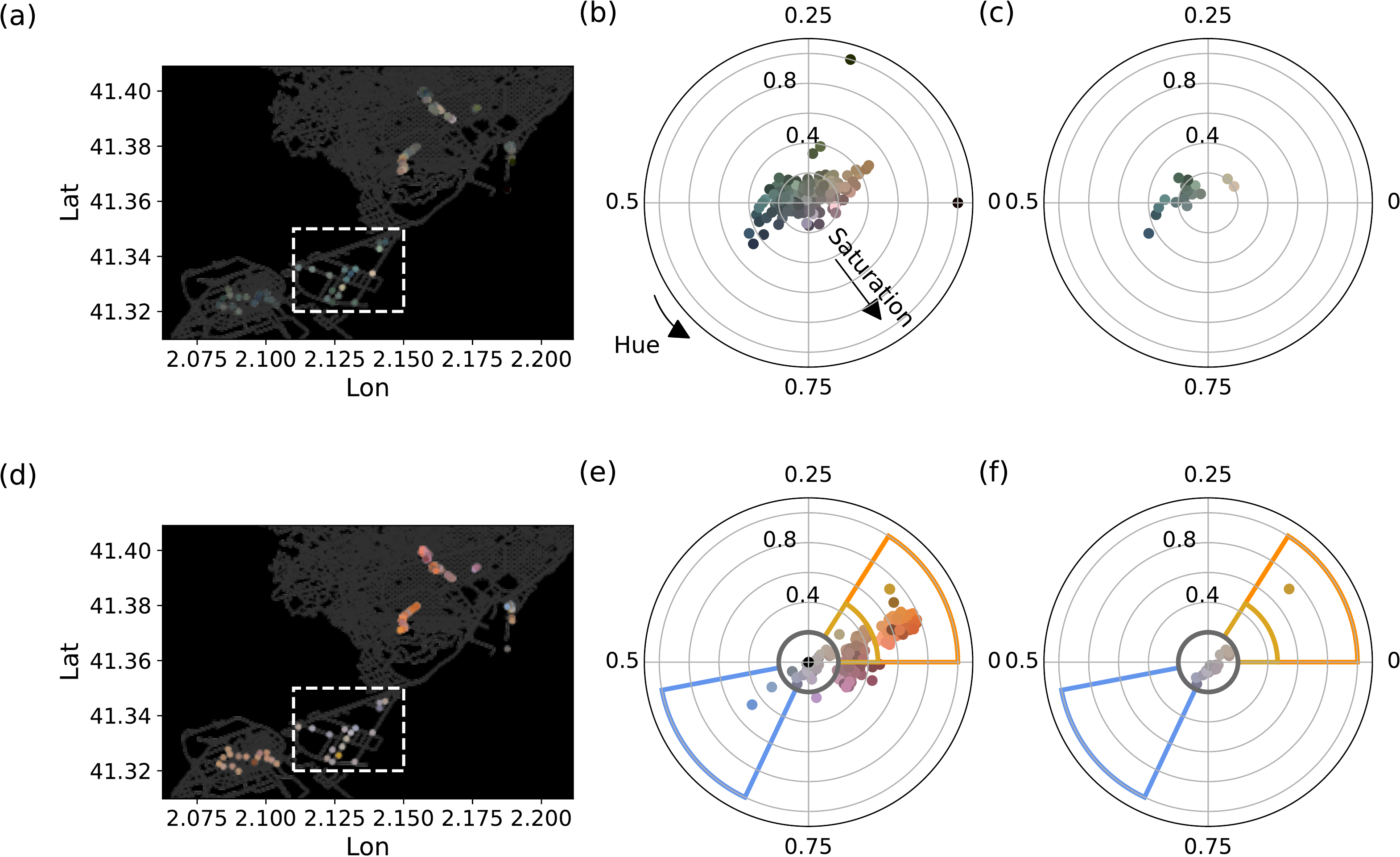}
	\caption{Analysis of an unlabelled dataset with the ISS (top row) and NightUp (bottom row). \textbf{(a,d)} Localization of the streetlights photographed for the unlabelled dataset. The color of the marker corresponds to the extracted color for that localization with the corresponding data source. \textbf{(b,e)} H-S scatter plot of the extracted colours from the previous unlabelled dataset; \textbf{(c,f)}. Same but for the boxed region in (a,d).} 
	\label{fig:comp_bcn}
\end{figure}

We finally study how our method (see Sec. \ref{sec:Castelldefels}) can classify unlabelled images according to their colour. To this end, in Fall 2020, we performed additional data acquisition in the cities of Barcelona and El Prat de Llobregat, for which we did not have any official database to compare with. Figure~\ref{fig:comp_bcn}(a, d) show the colours extracted from the ISS ISS062-E-102578 image and the images gathered with the NightUp app, respectively. The latter were taken by trained citizens which additionally reported the visual colour of each streetlight photographed. While in this case we don't have access to an official database of lamp types, we use the visual report of the trained citizens in order to validate the following analysis: in the Barcelona and El Prat de Llobregat urban areas, they report mainly yellowish lamps, while the photos in the industrial area of Barcelona were of whitish streetlights (boxed region in Figure~\ref{fig:comp_bcn}(a, d)). This allowed us to apply our method with more non-HPS lamps, that were scarce in Castelldefels, but abundant in the boxed region. Figure~\ref{fig:comp_bcn}(b,e) show the H-S diagram for the whole unlabelled dataset both for the ISS  and for the NightUp images, respectively. We can see how NightUp data are much more clustered than the ISS ones: we can distinguish various groups of colours, which, after inspection by the trained citizens, are directly related to the type of lamp that was pictured. To showcase this, we show in Figure~\ref{fig:comp_bcn}(c,f) an excerpt of the previous dataset only containing the images taken in the boxed region. The trained citizens reported that in that area there were only whitish lamps, except for one that was clearly yellow. This is reflected in Figure~\ref{fig:comp_bcn}(f), where one can see that our analysis classifies these images in the neutral region, clearly separated from the rest of images of the unlabelled dataset. Importantly, the automatic colour analysis matches the colours described by the humans experts, hence further highlighting the validity of the proposed method.

\section{Discussion}

The NightUp citizen science experiment aims at collecting information about the spatial distribution of colour of artificial lights at night, a powerful tool to understand light pollution, as the effects of light on the behaviour and health of humans, animals and plants varies enormously depending of the range of the visible spectrum considered. By means of a cross-platform mobile application, users were asked to take photos of streetlights. Then, we devised an algorithm to detect the lamps in the photos and extract their colour. With the geolocation of the NightUp data, we created a map that displays the colour of lamps in different regions. 
In order to validate the method before deploying it at large scale, we focused a first pilot NightUp campaign in the Castelldefels area, near Barcelona (Spain), for which we had an updated streetlight database provided by the city council, containing the geolocation and the lamp technology (e.g. high-pressure sodium or LED) for each streetlight in the city. With this information, we were able to test the NightUp experiment in various conditions, testing both the accuracy of the data acquisition as well as the colour extraction algorithm. In particular, we show that our method gives a estimation of the color of artificial light at night, sufficiently precise to distinguish warmer light sources from colder ones without a specific calibration of the device. Thanks to the comparison with the city council database, we identify the area of the H-S diagram associated to lamps emitting warm light. We define other four areas associated to different classes of lamp colours, and with it create an automatic classification algorithm of new photos, which we will use in future NightUp campaigns to automatically classify the photos from users. 

As final benchmark, and in order to directly test the method w.r.t state-of-the-art approaches, we compare our results with those obtained by analyzing satellite images taken from the ISS. The satellite approach has different benefits, like being able to characterize the colour of night illumination of large regions with very few, or even a single image shot. However, this comes with a smaller spatial resolution, that makes characterizing densely illuminated regions very hard. Moreover, the presence of obstacles or the fact that streetlights are pointed towards the ground heavily affect the colour analysis with this images. When compared to NightUp, we clearly see that colour extraction with the latter leads to clear distinctions between lamp colour categories, allowing to distinguish the HPS lamps from the other technologies, while the colours extracted from the ISS images can be hardly differentiated.

In general there is no other field work similar on concept to what NightUp propouse and only other light pollution programs like Globe at Night\cite{walker2008globe} and Nachtlichter similar\cite{weinberger2021value} and the ongoing Street Spectra project\cite{streetspectra} could reach similar outcomes with much more resources.

To summarize, NightUp allows for an efficient and accurate estimation of the colours of streetlights in densely populated regions. The information extracted from this analysis could be then used by local governments to optimize outdoor lighting and address light pollution problems in their regions as well as by scientist in the light pollution community to expand the studies on the effect of the colour of light. For example, as an important outlook for this project, one can use the colours extracted from the NightUp data to efficiently calibrate satellite images, as for example the one arising from the Visible and Infrared Scanner (VIRS), to improve the precision of ecological and economical parameters extracted from them. 

Furthermore, because of its minimal technical requirements and the simple storytelling, we believe that NightUp can potentially expand to a global scale with the support of successful engagement strategies, allowing scientists to access and then characterize vast regions of the world with very few measurements.

\section*{Authors contribution}
Conceptualization, Federica Beduini and Alejandro Sánchez De Miguel; Data curation, Gorka Muñoz-Gil, Alexandre Dauphin and Alejandro Sánchez De Miguel; Methodology, Gorka Muñoz-Gil, Alexandre Dauphin and Alejandro Sánchez De Miguel; Project administration, Federica Beduini; Software, Gorka Muñoz-Gil and Alexandre Dauphin; Validation, Alejandro Sánchez De Miguel; Visualization, Gorka Muñoz-Gil and Alexandre Dauphin; Writing – original draft, Gorka Muñoz-Gil, Alexandre Dauphin, Federica Beduini and Alejandro Sánchez De Miguel; Writing – review \& editing, Gorka Muñoz-Gil, Alexandre Dauphin, Federica Beduini and Alejandro Sánchez De Miguel.

\section*{Data availability}
The necessary code to reproduce the results shown in this paper, as well as a representative sample of the images analyzed can be found in the repository: \url{https://github.com/gorkamunoz/NightUp}

\acknowledgments{Thanks to the "Cities at Night" Program for the ISS images. The images from the ISS are courtesy of the Earth Science and Remote Sensing Unit, NASA Johnson Space Center. We acknowledge Scifabric for the design of the NightUp web application. Thanks to Lluís Torner and Silvia Carrasco for the institutional support and valuable discussions and advice. Thanks to David Paredes-Barato for his technical support and Lydia Sanmartí-Vila and Silvia Tognetti for participating in the data collection and engagement activities. We thank Castelldefels city council, that provided the streetlight database that allowed us to validate our data. Special thanks go to Alfonso López Borgoñoz, who helped us in different phases of the citizen science project. The Ramón Fernández Jurado library (especially its director Marta Granel), the Aula Senior project and the Col·legi Frangoal were crucial in helping us engaging with the people of Castelldefels. Last, but not least, all the anonymous citizens who contributed to NightUp deserve a special mention here because none of these results would have been possible without them.}

\section*{Funding}
This project has received funding from the European Union’s Horizon 2020 research and innovation programme under the Marie Skłodowska-Curie grant agreement No 847635 (UNA4CAREER). RALAN map project. This work was supported by the EMISSI@N project (NERC grant NE/P01156X/1). G.M-G. acknowledges support from the Austrian Science Fund (FWF) through SFB BeyondC F7102 and from Fundació LaCaixa. A.D. ERC AdG NOQIA; Ministerio de Ciencia y Innovation Agencia Estatal de Investigaciones (PGC2018-097027-B-I00/10.13039/501100011033,  CEX2019-000910-S/10.13039/501100011033, Plan National FIDEUA PID2019-106901GB-I00, FPI, QUANTERA MAQS PCI2019-111828-2, QUANTERA DYNAMITE PCI2022-132919,  Proyectos de I+D+I “Retos Colaboración” QUSPIN RTC2019-007196-7); European Union NextGenerationEU (PRTR);  Fundació Cellex; Fundació Mir-Puig; Generalitat de Catalunya (European Social Fund FEDER and CERCA program (AGAUR Grant No. 2017 SGR 134, QuantumCAT \ U16-011424, co-funded by ERDF Operational Program of Catalonia 2014-2020); Barcelona Supercomputing Center MareNostrum (FI-2022-1-0042); EU Horizon 2020 FET-OPEN OPTOlogic (Grant No 899794); National Science Centre, Poland (Symfonia Grant No. 2016/20/W/ST4/00314); European Union’s Horizon 2020 research and innovation programme under the Marie-Skłodowska-Curie grant agreement No 101029393 (STREDCH) and No 847648  (“La Caixa” Junior Leaders fellowships ID100010434: LCF/BQ/PI19/11690013, LCF/BQ/PI20/11760031,  LCF/BQ/PR20/11770012, LCF/BQ/PR21/11840013).  A.D. further acknowledges the financial support from a fellowship granted by la Caixa Foundation (ID 100010434, fellowship code LCF/BQ/PR20/11770012). Project cofinanced by the Diputació de Barcelona through the BiblioLab program (21296).

\section*{Conflict of Interests}
A. Sánchez de Miguel discloses that he does consulting work sporadically for Savestars Consulting S.L. and is a member of the board of Cel Fosc. The authors are not aware of any affiliations, memberships, funding, or financial holdings that might be perceived as affecting the objectivity of this article.

\appendix
\section[\appendixname~\thesection]{ISS image transformation procedure}
\label{app:iss_transformation}

In this paper we use the ISS image ISS062-E-102578, which we first process following the procedure proposed in Ref.\cite{de2021colour} using synthetic photometry from high resolution spectra. As we stated in section \ref{sec:ISS_sub}, the ISS image values are in physical units (radiance values), whereas the NightUp ones are in the JPEG format, with arbitrary units.

In order to transform from radiance values to the HS values used in this paper, we use the following sequence of manipulations, which effectively generate image values comparable to the JPEG ones. The ISS plots do not use the same CCT, contrast and saturation adjustments of the NightUp images, so they are not directly comparable, although there is a homomorphic transformation between them. In order to compare the different datasets, we set the center of the HS plot on known areas of pure white sources: this allows us to compare the two datasets, but it may also create a rotation on the HS plane and stretching effect. We then rescale the three channels following $x=(x-x_{\min)}/x_{\max}$ with $r_{\min},g_{\min},b_{\min}= 0.01, 0.01, 0.003$ and  $r_{\max},g_{\max},b_{\max}=0.681,0.603,0.558$. The $r_{\min},g_{\min},b_{\min}$ values have been selected from the background of the delta of the Llobregat river, a non-illuminated area in the ISS image. With this operation, some negative values that do not have physical meaning may appear: they come from the noise distribution of the background. In order to avoid these artifacts, we set all the negative values to 0. We then apply the so-called gamma correction $x\rightarrow x^{\gamma_x}$, with $\gamma_r,\gamma_g,\gamma_b =1/1.8, 1/2.8, 1/3.9$. Finally, we transform the new RGB amplitudes to HSV~\cite{burger2009}. As the JPEG images are in arbitrary units, we can choose the values $x_{\max}$ and $\gamma_x$ ad hoc, so that the visual comparison between the two data sets becomes easier.

\bibliographystyle{apsrev4-2}
\bibliography{bibliography}

\end{document}